# Security Protocols in a Nutshell

Mohsen Toorani

Department of Informatics

University of Bergen, Norway

mohsen.toorani@uib.no

**Abstract**

Security protocols are building blocks in secure communications. They deploy some security mechanisms to provide certain security services. Security protocols are considered abstract when analyzed, but they can have extra vulnerabilities when implemented. This manuscript provides a holistic study on security protocols. It reviews foundations of security protocols, taxonomy of attacks on security protocols and their implementations, and different methods and models for security analysis of protocols. Specifically, it clarifies differences between information-theoretic and computational security, and computational and symbolic models. Furthermore, a survey on computational security models for authenticated key exchange (AKE) and password-authenticated key exchange (PAKE) protocols, as the most important and well-studied type of security protocols, is provided.

# Contents

# 1 Introduction

The security is about protection of *assets* from various *threats*, posed by certain inherent *vulnerabilities*. A threat is a potential violation of the security, while an attack is a threat that is carried out. Security processes deal with selection and implementation of *security controls* (also called countermeasures) which help to reduce the *risk* posed by the vulnerabilities [1].

Security protocols are a hot topic in information and cyber security. They are building blocks in secure communications. *Information security* and *cyber security* are used interchangeably in the literature, but they are not the same. Figure 1 depicts the relationship among the cyber security, information security, and information and communication technology (ICT) security.

A *protocol* is a set of rules or conventions which defines an exchange of messages between a set of two or more partners. Protocol actors or partners can be users, processes or machines, and are generally referred to as *principals* [2]. Each principal is usually associated with an identity, and plays a protocol role. A protocol role defines the rules that a principal playing that role should obey.

A *security protocol* is an abstract or concrete protocol which performs some security-related functions. The goal is to provide certain desirable security services and protection against attacks. A security protocol applies some security mechanisms, maybe as sequences of cryptographic primitives. There are dishonest actors and attackers that are not constrained to follow the protocol rules, and try to prevent protocol executions from reaching their security goals. An attacker has many ways to deal with a protocol and may execute multiple sessions of a protocol concurrently. The problem of protocol composition is also a challenge [3]. Larger protocols are often composed of some small protocols. Messages generated by a run of a protocol might be useful for an attack on another protocol [4].

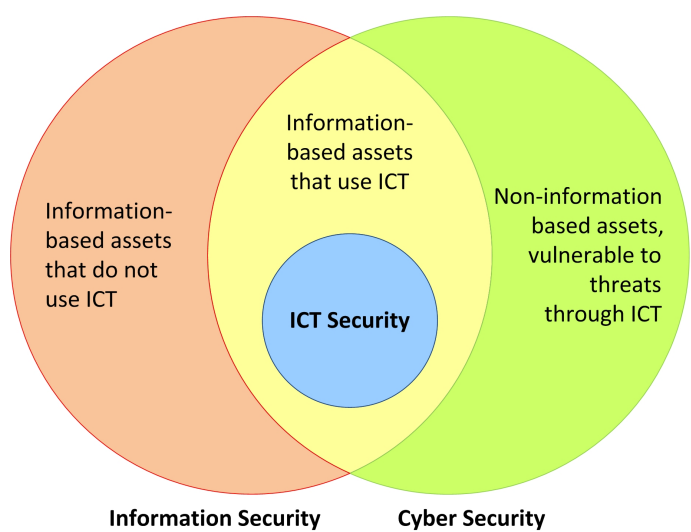

Figure 1: The relationship between information security and cyber security [1]

## 1.1 Security attacks

Threats in information security can be divided into four broad classes: disclosure or unauthorized access to information; deception or acceptance of false data; disruption, interruption or prevention of correct operation; and usurpation or unauthorized control of some part of a system [5]. Attacks can also be classified into passive and active attacks [6]. *Snooping* or *eavesdropping* is a kind of disclosure and unauthorized interception of information. It is a passive attack, and can be either release of messages contents, traffic analysis, or browsing through files or system information. *Masquerading* or *spoofing* takes place when an entity pretends to be another entity. It can be considered as a type of deception and usurpation. Unauthorized *modification of information* is an active attack which can be deception, or disruption and usurpation. *Repudiation of origin* where an entity denies sending or creating something, and *denial of receipt* in which an entity denies receiving some information or messages can be considered as type of deception. *Delay* and *denial of service (DoS)* is a temporary and long-term inhibition of a service, respectively. They can be considered as usurpation, although they can play a supporting role in deception. They may result from direct attacks or from other problems that are not related to security.

## 1.2 Security services

Confidentiality, integrity, and availability are basic security services in information and cyber security. They are sometimes referred to as the *CIA triad*. The interpretations of these aspects vary according to the contexts in which they arise.

*Confidentiality* is the concealment of information or resources [5]. Information has confidentiality when it is protected from disclosure or exposure to unauthorized individuals or systems [7]. *Privacy* is sometimes defined through and included in the confidentiality [8].

*Integrity* refers to trustworthiness of data or resources. It preserves data as complete, and uncorrupted; and prevents improper or unauthorized change to the data. *Authentication* and *non-repudiation* are sometimes defined through integrity [8]. Integrity may refer to *data integrity* which considers content of data, and *origin integrity* (or authentication) which considers source of data. The integrity is threatened when data is exposed to corruption, damage, destruction, or other disruption of its authenticity [7].

*Availability* ensures timely and reliable access to and use of information [8]. It enables authorized entities to access information and systems, and can be considered as an aspect of *reliability*. DoS attacks aim to defeat the availability.

It has been argued that the CIA triad does not provide a complete set of required security goals [7,9]. *Authenticity* and *accountability* are two aspects that are considered as additional requirements [6], although the authenticity are sometimes defined through the integrity in the CIA triad. *Accountability* enables tracing actions of an entity. It supports non-repudiation, deterrence, fault isolation, intrusion detection and prevention, after-action recovery and legal action [10]. Recently, the following list has been proposed for covering different aspects of information assurance and security: Confidentiality, integrity, availability, accountability, auditability, authenticity/trustworthiness, non-repudiation, and privacy [9].

### 1.3 Security mechanisms

A security mechanism is a process or device that is designed to prevent, detect, or recover from a security attack [6]. Prevention mechanisms may include use of cryptographic schemes. A *cryptographic scheme* is a suite of related cryptographic algorithms and/or protocols. A *cryptographic algorithm* is a well-defined transformation, which on a given input value produces an output value, and provides certain security objectives. A *cryptographic protocol* is a distributed algorithm that describes precisely the interactions between two or more entities, in order to achieve certain security objectives. Detection mechanisms aim to determine that an attack is underway, or has occurred. Recovery mechanisms may require resumption of operation to stop an attack and repair damages, or the system may continue functionality or keeping essential functionalities while an attack is underway. Recovery may also include retaliation by attacking the attacker's system or taking legal steps to prevent an attacker from repeating an attack [5].

## 2 Taxonomy of attacks

In this section, we briefly review some attacks on protocols, encryption schemes, and their implementations. Attacks on other cryptographic primitives are beyond our discussion. By implementation attacks, we mean attacks which use information, which is leaked by a cryptographic primitive or by its particular usage within a security protocol. For example, measuring power consumption or the time taken to encrypt the same message with different secret keys can infer some information about the secret keys.

### 2.1 Attacks on security protocols

An attack on a protocol is defined and accomplished according to security objectives or security requirements of a protocol, or the security model wherein the security of a protocol is proved. An attack occurs when any goal of the protocol is violated. Attacks on security protocols can be broadly divided into passive and active attacks. It is also possible to categorize attacks on security protocols based on the weaknesses they exploit [11–13].

In passive attacks, an attacker just intercepts, records or analyzes protocol messages; while in active attacks, an attacker also alters, deletes, redirects, reorders, forges, and inserts new protocol messages into the conversation. *Eavesdropping* is a passive attack where an adversary captures the messages sent in the protocol execution. It aims to defeat the confidentiality or privacy. It is also a prerequisite for many sophisticated attacks. In the computational security settings, an adversary is allowed to eavesdrop on messages. This attack is very practical, because many communication systems are prone to this attack [14]. Encryption of messages is a countermeasure to this attack. *Modification of messages* is an active attack where an adversary modifies the information sent in the protocol run. It aims to defeat the integrity. This attack can be thwarted by integrity mechanisms. However, encryption of messages is not by itself sufficient for providing the required integrity properties [15].

In this section, we present a non-exhaustive list of standard attacks on protocols [16–19]. They are the most common types of attacks, based on practical scenarios whereby an attacker can cause a protocol failure. The list is incomplete because there are unlimited ways for an adversary to interact with one or more (e.g. parallel) protocol executions. The following list does not include attacks based on flaws in the hardware or software implementations. Security against a list of attacks does not guarantee the security of a protocol, but one would expect that a new protocol does not inherit failures from previous protocol designs. Furthermore, a security model is expected to address most important attacks that take place in reality. This is a good measurement for evaluation of security models: If a security model does not allow an adversary to perform attacks that can take place in reality, the security proofs will be useless, because there will be practical scenarios for breaking the protocol, which are not captured by the security model.

- *Impersonation attack*: It is an active attack which aims to defeat the authenticity. In this attack, an adversary tries to impersonate one or more entities. According to the adversarial model in the corresponding security model, the impersonation attack may have weaker variants in authenticated key exchange (AKE) or password-based authenticated key exchange (PAKE) protocols. *Key compromise impersonation* (KCI) attack and *ephemeral key compromise impersonation* attack are weaker variants of the impersonation attack in AKE protocols, which require knowledge of a static private key and an ephemeral private key (a random number), respectively. The goal in such variants is to impersonate another entity to the compromised entity [20].

- *Man-in-the-middle (MITM) attack*: It is a variant of the impersonation attack, where an adversary resides between two entities, and convincingly impersonates both victims. Practical examples include MITM attack on the GSM cellular network [21], the HTTPS protocol [22], and the EMV (Europay, MasterCard and Visa) protocol [23]. MITM is feasible when a protocol lacks (mutual) authentication.

- *Unknown key-share (UKS) attack*: It is a variant of the impersonation attack in AKE protocols. In a UKS attack, two entities share a session key, but they have different views about the identity of their peer [24]. The UKS attack is feasible when a key exchange protocol fails to provide an authenticated binding between the session key and identifiers of the honest entities [25]. Typically, there are two kinds of UKS attacks [26, 27]: In the first type, which is referred to as a *Public key substitution UKS attack*, an adversary registers another entity's public key as its own public key. In the second type UKS attack, the adversary has a valid public-private key certified by the CA, and tries to perform a UKS attack.

- *Replay attack*: It is an active attack in which an adversary interferes with a protocol run by insertion of some messages from previous protocol runs or parallel sessions. It can be considered as a combination of eavesdropping and modification attacks. A protocol is vulnerable to the replay attack if it fails to provide *freshness*. The freshness can be provided using timestamps, nonces or session tokens, and counters [28]. A taxonomy of replay attacks on cryptographic protocols is provided in [29]: Messages can be replayed from inside or outside the current run of the protocol. Messages can be directed to intended principal by delay (*straight replay*), or they can be directed to other entities instead of the intended recipient (*deflection*). In deflection attacks, the messages can be sent back to the sender (*reflection*) or sent to a third party.

  The *Reflection attack*, which is a special case of the replay attack, may be considered as an *oracle attack*. In an oracle attack, an adversary uses a protocol actor as an oracle to get some information that the adversary cannot generate on its own. The adversary may use this information to forge new messages that are injected to another parallel protocol session.

- *Preplay attack*: In a preplay attack, the adversary prepares for the attack in advance by simulating a protocol execution and performing a set of operations. The adversary performs the real attack later when it is likely to carry out the same series of operations as in the simulation. The preplay attack is feasible when challenge is predictable in challenge-response protocols [28, 30]. An example of preplay attack is cloning the EMV cards [31–33].

- *Denial of service (DoS) attack*: DoS attacks refer to a broad class of attacks that target availability of the systems [34]. In terms of protocols, they refer to an attack whereby an adversary prevents legitimate entities from completing a protocol. In practice, they may take place against servers that interact with many clients. An attacker may use up the computational resources of the intended server (*resource depletion* attack), or exceeds the number of allowed connections to the server (*connection depletion* attack). It is impossible to prevent DoS attacks completely, but it is possible to reduce their impacts. Protocols that postpone authentications to the end of the protocol are much more vulnerable to the DoS attack than protocols that take care of authentication at earlier stages. Meadows [35] suggested the principle of gradual authentication, and developed a formal framework based on *fail-stop* protocols [36]. Each message in a protocol must be authenticated, but authentications can be weak in the beginning, and become strong in subsequent protocol steps. A protocol is fail-stop if any bogus message can be detected, and the protocol halts upon detection. Other strategies proposed for strengthening protocols against DoS attacks include counterbalancing memory expenditure and counterbalancing computational expenditure [37]. Counterbalancing memory expenditure can be achieved by using *cookies*. Cookies were first introduced in the Photuris session-key management protocol [38], and subsequently extended for resisting SYN flooding DoS attacks [39]. They are a common way for implementing stateless connections [40] and allow servers to remain stateless during execution of the protocol. A common way of counterbalancing computational expenditure is via the use of *client puzzles* [41, 42], which increases the amount of computational resources required for mounting a DoS attack [37].

- *Type flaw attacks*: In type flaw attacks, an attacker uses the absence of proper message type checking. The attacker sends a message of different type than what is expected. The victim entity fails to detect the type mismatch, and misinterprets the message content or behave in an unexpected way. As an example, the Otway-Rees key transport protocol [43] is vulnerable to this attack [19]. Countermeasures to the type flaw attack include changing the order of message elements in next uses of the same message, and ensuring that each encryption key is used once. Other methods may include an authenticated message number in each message, or an authenticated type field with each field [44].

- *Cryptanalysis*: In security protocols, cryptographic primitives are considered abstract, and secure against attacks. However, there is an exception when the key is known to be weak. Those situations should not expose verifiers or evidences that can be used for deducting the key. For example, the PAKE protocols use a human-memorable password, of low entropy for authentication and establishing a strong cryptographic key. It is crucial for PAKE protocols to resist against the following attacks:

    - *Offline dictionary attack*: In an offline dictionary attack which is a passive attack, the adversary eavesdrops communication between two honest entities,

and obtains a verifier that can be used for extracting the password using a dictionary of most probable passwords. The adversary applies each password from the dictionary to the obtained verifier until she finds the correct password that satisfies the verifier equation.

- *Online dictionary attack*: In an online dictionary attack, the adversary uses a dictionary of most probable passwords, but obtains the verifier through an online interaction with the target entity. As a countermeasure, servers usually lock the account of a user after number of unsuccessful trials. There is a more complicated kind of this attack, called an *undetectable online dictionary attack* [45] that should be thwarted by the PAKE protocols.

- *Chosen protocol attack*: In a chosen protocol attack, a new protocol is designed to interact with an existing protocol, and to create a security hole. This attack is based on a scenario for protocol interactions where a key is used for multiple applications e.g. smart cards. Several examples of this attack are provided in [46]. This attack is sometimes referred to as the *multi-protocol* attack [47].

- *Internal action flaws*: A group of attacks are based on absence of some operations that are crucial to guarantee a security property. An example is absence of a checking on the message received in the third phase of the *Three Pass Protocol* [48]. Other examples include:

  - *Small-subgroup attack*: In protocols based on discrete logarithm problem, a small-subgroup attack [49–51] would be feasible if ephemeral or static public keys are not validated to be of prime order. An attacker can then select a group element of small order to enforce the session key or a verifier to lie in a group of small order. This can help an adversary by reducing the number of possibilities in a brute-force attack. This attack can be used for extracting a shared session key, password, or even the private key of an entity. A countermeasure is to check that the received elements reside in the correct group.

  - *Invalid-curve attack*: This attack is a variant of the small-subgroup attack on elliptic curve-based protocols. In an invalid-curve attack, an adversary selects a point of small order on an invalid-curve, and tries to extract the private key of an entity [52]. The attack would be feasible if an EC-based protocol does not consider validation of static or ephemeral public keys [53, 54]. There are other known attacks on elliptic curve-based schemes that will be feasible if domain parameters of elliptic curves do not satisfy some requirements [55, 56].

## 2.2 Attacks on encryption schemes

A naive way to attack an encryption scheme is through the *brute-force* or *exhaustive key search* attack, where an attacker tries all the possible keys in the keyspace on a pair of plaintext-ciphertext, until he finds the key. Excluding information-theoretically secure schemes, any encryption scheme in the computational setting is vulnerable to this attack unless the keyspace is large enough that makes the attack computationally infeasible. Attacks on encryption schemes can be categorized into the following attack models. The objective of an attacker is to systematically recover plaintext from ciphertext, or to deduce the key [16].

- In a *ciphertext-only attack*, the adversary has only the ciphertext. An encryption scheme is completely insecure if it is vulnerable to this attack.

- In a *known-plaintext attack*, the adversary also has a quantity of plaintext and the corresponding ciphertext.

- In a *chosen-plaintext attack* (CPA), the adversary selects plaintext, and is then given the corresponding ciphertext. The adversary uses the deduced information to recover the corresponding plaintext of a previously unseen ciphertext. Public key encryption schemes are an example where an adversary can encrypt any message of her choice under public key of the victim entity.
  The *adaptive chosen-plaintext attack* (CPA2) is a CPA attack in which the choice of plaintext by the adversary may depend on the ciphertext created in previous encryptions.

- In a *chosen-ciphertext attack* (CCA), an adversary is able to decrypt arbitrary ciphertexts, e.g. by access to a decryption equipment with a securely embedded decryption key. The goal is to deduce plaintext from previously unseen ciphertext. CCA has two special variants:
  In *non-adaptive chosen ciphertext attack* (CCA1) [57], which is also called *lunchtime* or *midnight* attack, an adversary can only have access to the system for limited time or limited number of plaintext-ciphertext pairs. The attack is called non-adaptive because an adversary cannot adapt her queries to the decryption oracle according to the challenge ciphertext. In CCA1, the challenge ciphertext is given after the adversary's ability to make chosen ciphertext queries has expired. However, an adversary may make adaptive chosen ciphertext queries before the challenge ciphertext is given. The term lunchtime refers to the idea of having accessing to the decryption oracle during a lunch break.
  In *adaptive chosen-ciphertext attack* (CCA2) [58], which is stronger than CCA1, an adversary has access to the decryption oracle even after having the challenge ciphertext. In CCA2, an adversary's queries to the decryption oracle may depend on the challenge ciphertext, but the adversary may not ask for the decryption of the challenge ciphertext itself.

Most of the above attacks may apply to digital signature schemes and message authentication codes (MAC) where the objective of an attacker is to forge messages or MACs. Based on the above attack models, different techniques of cryptanalysis have been introduced in the literature. Most widely used techniques for cryptanalysis of symmetric-key encryption schemes are differential cryptanalysis [59, 60], linear cryptanalysis [61] and algebraic cryptanalysis [62]. Other techniques include combined attacks [63], meet-in-the-middle attack [64], integral cryptanalysis [65], related-key attack [66], and distinguishing attack [67].

Asymmetric-key encryption schemes are built on intractability of some hard problems. Hard problems that are used in public key cryptography include integer factorization, discrete logarithm problem (DLP) in appropriate groups such as multiplicative groups of finite fields or additive groups of elliptic curves over finite fields, knapsacks and lattice problems, coding problems, and multivariate polynomial equations over small finite fields. Cryptanalysis of an asymmetric-key encryption scheme may rely on incorrect assumptions about intractability of the hard problem, or design failures or problems due to implementation. In order of increasing strength, mostly used attack models on public key encryption schemes are CPA, CCA1 [57], and CCA2 [58]. *Parallel attacks* [68] can be considered as variants to those attack models. Parallel attacks was developed to connect notions of *non-malleability* with *indistinguishability*. In parallel attacks, an adversary may ask for the decryption of some ciphertexts after getting the challenge ciphertext, but cannot ask for decryption of the challenge ciphertext. However, queries may not depend on the result of each other, and may be viewed as being processed in parallel, hence yielding the name parallel attack. Some attack techniques have also been proposed for probabilistic public key cryptosystems. This includes ciphertext-verification attack [69] or plaintext-checking attack [70], and reaction attack [71].

Several notions of security have been introduced for encryption schemes. This includes

- *Indistinguishability* (IND) [72] which formalizes an adversary's inability to learn any information about the plaintext underlying a challenge ciphertext.

- *Non-malleability* (NM) [73–76] which formalizes an adversary's inability to transform a given ciphertext into a different ciphertext so that their according plaintexts are "meaningfully related". Non-malleability was originally defined as simulation-based NM by Dolev et al. [73–75]. Bellare et al. [77, 78] introduced comparison-based NM later, but it was shown [68] that both definitions specify an equivalent security model, and NM can be reduced to a certain model of IND with a more practical formalisation.

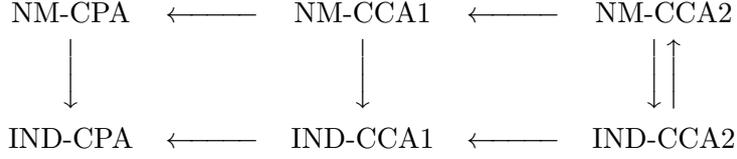

Figure 2: Relations among notions of security for public-key encryption schemes [77]

- *Plaintext awareness* [77, 79] which formalizes an adversary's inability to create a ciphertext without knowing the underlying messages.

Indistinguishability is an important property for preserving the confidentiality. However, in some cases, it may imply other security services like integrity, which is somehow related to non-malleability [80]. Figure 2 depicts relations among notions of indistinguishability and non-malleability for public-key encryption schemes under CPA, CCA1, and CCA2 attacks [77]. Arrows denote implications. For example, if an encryption scheme is NM-CCA2 secure, it is also NM-CCA1 secure. However, if an encryption scheme is NM-CCA1 secure, it might be broken in the NM-CCA2 sense. IND-CCA2 implies all other notions. Notions of {IND-CPA, IND-CCA1, IND-CCA2} and {NM-CPA, NM-CCA1, NM-CCA2} can be defined as follows.

**Definition 1. *IND-CPA, IND-CCA1, IND-CCA2*.** Let $\Pi = (Gen, Enc, Dec)$ denote a public key encryption scheme, and let $\mathcal{A} = (\mathcal{A}_1, \mathcal{A}_2)$ denote an adversary with two sub-algorithms. For attack $atk \in \{cpa, cca1, cca2\}$ and security parameter $n \in \mathbb{N}$, the adversary's probability of success is defined as $\mathbf{Adv}_{\mathcal{A},\Pi}^{ind-atk}(n) = Pr[\mathbf{Exp}_{\mathcal{A},\Pi}^{ind-atk-1}(n) = 1] - Pr[\mathbf{Exp}_{\mathcal{A},\Pi}^{ind-atk-0}(n) = 1]$ in which $b \in \{0, 1\}$, the experiment $\mathbf{Exp}_{\mathcal{A},\Pi}^{ind-atk-b}(n) = b'$ is defined as

$$
\begin{aligned}
(pk, sk) &\leftarrow Gen(1^n) \\
(m_0, m_1, s) &\leftarrow \mathcal{A}_1^{\mathcal{O}_1}(pk) \\
b &\in_R \{0, 1\} \\
c &\leftarrow Enc_{pk}(m_b) \\
b' &\leftarrow \mathcal{A}_2^{\mathcal{O}_2}(m_0, m_1, s, c) \\
&\text{return } b'
\end{aligned}
$$

where:

$$
\begin{aligned}
&\textit{For atk} = cpa, \quad \mathcal{O}_1(.) = \varepsilon \quad \textit{and} \quad \mathcal{O}_2(.) = \varepsilon \\
&\textit{For atk} = cca1, \quad \mathcal{O}_1(.) = \mathcal{O}_{Dec}(.) \quad \textit{and} \quad \mathcal{O}_2(.) = \varepsilon \\
&\textit{For atk} = cca2, \quad \mathcal{O}_1(.) = \mathcal{O}_{Dec}(.) \quad \textit{and} \quad \mathcal{O}_2(.) = \mathcal{O}_{Dec}(.)
\end{aligned}
$$

An encryption scheme is secure in the sense of IND-ATK if $\boldsymbol{Adv}_{\mathcal{A},\Pi}^{ind-atk}(.)$ is negligible in n [77].

**Definition 2.** ***NM-CPA, NM-CCA1, NM-CCA2.*** *Let* $\Pi = (Gen, Enc, Dec)$ *denote a public key encryption scheme, and let* $\mathcal{A} = (\mathcal{A}_1, \mathcal{A}_2)$ *denote an adversary with two sub-algorithms. For attack* $atk \in \{cpa, cca1, cca2\}$ *and security parameter* $n \in \mathbb{N}$, *the adversary's probability of success is defined as*
$\boldsymbol{Adv}_{\mathcal{A},\Pi}^{nm-atk}(n) = Pr[\boldsymbol{Exp}_{\mathcal{A},\Pi}^{nm-atk-1}(n) = 1] - Pr[\boldsymbol{Exp}_{\mathcal{A},\Pi}^{nm-atk-0}(n) = 1]$ *in which for* $b \in \{0, 1\}$, *the experiment* $\boldsymbol{Exp}_{\mathcal{A},\Pi}^{nm-atk-b}(n) = b'$ *is defined as*

$$
\begin{aligned}
(pk, sk) &\leftarrow Gen(1^n) \\
(m_0, m_1, s) &\leftarrow \mathcal{A}_1^{\mathcal{O}_1}(pk) \\
c &\leftarrow Enc_{pk}(m_1) \\
(R, \mathbf{c}) &\leftarrow \mathcal{A}_2^{\mathcal{O}_2}(m_0, m_1, s, c) \\
\mathbf{m} &\leftarrow Dec_{sk}(\mathbf{c}) \\
\text{If } c \notin \mathbf{c} \wedge \perp \notin \mathbf{m} \wedge R(m_b, \mathbf{m}) \quad \text{then} &\quad b' \leftarrow 1 \\
\text{Otherwise} &\quad b' \leftarrow 0 \\
\text{return } b'&
\end{aligned}
$$

where $\mathbf{m}$ and $\mathbf{c}$ denote vectors of plaintexts and ciphertexts, $\perp$ denotes the output of decryption, $R(.)$ represents the notion of being "meaningfully related", and

$$
\begin{aligned}
\text{For } atk = cpa, \quad &\mathcal{O}_1(.) = \varepsilon \quad \text{and} \quad \mathcal{O}_2(.) = \varepsilon \\
\text{For } atk = cca1, \quad &\mathcal{O}_1(.) = \mathcal{O}_{Dec}(.) \quad \text{and} \quad \mathcal{O}_2(.) = \varepsilon \\
\text{For } atk = cca2, \quad &\mathcal{O}_1(.) = \mathcal{O}_{Dec}(.) \quad \text{and} \quad \mathcal{O}_2(.) = \mathcal{O}_{Dec}(.)
\end{aligned}
$$

An encryption scheme is secure in the sense of NM-ATK if $\boldsymbol{Adv}_{\mathcal{A},\Pi}^{nm-atk}(.)$ is negligible in n [77].

Notions of security in symmetric-key encryption schemes can also be modeled as games, played with an adversary [81]. Relations among notions of security for symmetric encryption have been discussed in [82]. Relations among notions of plaintext awareness for encryption schemes have been considered in [83].

## 2.3 Attacks on implementations

Cryptographic primitives and protocols are conventionally considered abstract when they are designed. This mathematical abstraction is useful in the study, but it does not capture the whole scenario that can happen in practice. Cryptographic algorithms are always implemented in software or hardware on physical devices that are influenced by the environment. An attacker may not need to directly take on the computational complexity of breaking the algorithms in order to derive the plaintext or the key. Information obtained by observing the computations or communication of a concrete implementation may help a lot in cryptanalysis, and can greatly decrease the computational complexity of breaking a cryptosystem. In the following, we briefly review some attacks on implementation of cryptosystems. There are other sources of errors that can be used for attacks on implementation, but we do not consider them. This includes logical flaws and bugs in software implementation, or any divergence between the protocol specification and its implementation.

- *Side-channel attacks*: Side-channel attacks [84–86] are low cost, realistic, and usually considered as the most dangerous type of physical attacks. Computing devices leak information not just through input-output interaction, but through physical characteristics of computation such as power consumption, timing, or electromagnetic radiation. Such information leakage can break many cryptosystems in common use, and are feasible when an adversary has access to the device, as it is often the case for devices such as smartcards, TPM chips, mobile phones and laptops [87]. Side-channel attacks can be divided into passive and active attacks. In passive side-channel attacks, an adversary does not interfere with the operation of the target system; while in active side-channel attacks (or fault attacks), the adversary has some influence on the behavior of the target system. Side-channel attacks can also be classified as invasive, non-invasive, and semi-invasive [88]. Invasive attacks require depacking to get direct access to the internal components of the device. Non-invasive attacks only exploit externally available information. Semi-invasive attacks include access to the device without damaging the passivation layer or having electrical contact with unauthorized surface. The following list provides an incomplete list of side-channel attacks. It is also possible to exploit combination of side-channel attacks. *Multi-channel attacks* simultaneously utilize multiple side-channels such as power and electromagnetic analysis [89]. Known side-channel attacks include:

    - *Timing attack*: In timing attack [90–92], some information about the key or secret parameter is deduced from the running time of a cryptographic algorithm or device. Timing analysis of keystroke may also help in finding the password [93].
    - *Power analysis attack*: In power analysis attacks [84], valuable information about operations or parameters are obtained by observing the power consumption of a cryptographic device or module. Power analysis attacks can be divided into simple power analysis (SPA), where the power consump-

tion measurements are directly interpreted, and differential power analysis (DPA), where statistical functions are applied to the power consumption measurements [94].

- *Electromagnetic analysis attack*: In electromagnetic analysis attack [95,96], some information is obtained by measuring the electromagnetic fields radiated from the device. This attack may also be used for detecting keystroke on a keyboard [97].

- *Acoustic cryptanalysis*: Acoustic emanations can be considered as a source of information for side-channel attacks. Examples include extracting 4096-bit RSA keys from the GnuPG software using the sound generated by the computer during decryption of some chosen ciphertexts [98], or recognizing the key being pressed by acoustic emanation of the keyboard [99, 100].

- *Padding oracle attacks*: The Padding oracle attack relies on having a *padding oracle* as a side-channel, which determines whether a message is correctly padded or not. In many standards, messages are pre-formatted before encryption, and the recipient sends an error message if the message is not correctly formatted [101]. This attack can be generalized on block ciphers in the CBC mode [102], and is feasible on RSA-OAEP PKCS#1 v.2.0 [103] and EME-OAEP PKCS#1 v.2.1 [104], web application frameworks [105, 106], and cryptographic devices [107].

- *Memory attacks*: Side-channel information from CPU cache [108, 109] and DRAM [110, 111] can be used for cryptanalysis of software-implemented ciphers. A CPU cache resides between CPU and the main memory to speed up the run-time. Cache-based attacks use measurement of the delay caused by a cache miss, which occurs when the CPU accesses data that were not stored on the cache, and has been used for cryptanalysis of ciphers including DES [108] and AES [109]. DRAM cells also retain their state for long intervals even if the computer is unplugged, which can deduce useful information about the secret key, and has been used against DES, AES, RSA and encrypted disks [110, 111].

- *Fault injection attacks*: Fault injection attacks are active counterpart of side-channel attacks, where an attacker gains information about internal states of the algorithm by provoking errors in the computation, and comparing the correct and erroneous output. The fault can be *permanent* which damages the cryptographic device irreversible, or it can be *transient*. The fault may be injected by change in voltage, clock frequency or temperature, or using light, x-ray and microwave radiation, or eddy current caused by magnetic fields. A survey on techniques and countermeasure is available in [112].

Proposed countermeasures to the side-channel attacks include ad-hoc solutions to the implementation which offer only partial remedy, and theoretical solutions which formally address the problem. *Leakage resilient cryptography* [113, 114] is an active area of research which concerns computing in presence of information leakage, and considers mathematical solutions to address side-channel attacks. A leakage-resilient cryptosystem remains secure even if arbitrary, yet bounded, information about the secret key and possibly other internal state information is leaked to an adversary [115].

- *Traffic analysis*: By traffic analysis, an adversary may deduce important information from patterns in communication. Patterns may include timing, delay, length, conversation partners or frequency. Such information may thwart the privacy or anonymity [116], or may speed up the process of finding a secret parameter [93]. An example is statistical analysis of frequency and timing of network packets for interactive sessions of the SSH protocol, which makes finding users' passwords fifty times faster than a brute-force attack [93]. The SSH protocol sends each keystroke in a separate packet during interactive sessions, which leaks the inter-keystroke timing of users' typing. Other examples include statistical disclosure or intersection attacks on anonymity systems [117], and website fingerprinting attacks [118].

## 3 Security models

Different approaches and methodologies have been considered for security evaluation in literature. According to Menezes et al. [16], models for evaluating security can be classified into five models: Unconditional security, complexity-theoretic security, provable security, computational security, and ad hoc (heuristic) security; along with the computational, provable and ad hoc methodologies are most practical, although the ad hoc security is usually prone to failure.

- **Unconditional security**: The most stringent measure where an adversary is assumed to have unlimited resources, but should not gain enough information to defeat the system. Unconditional security of encryption schemes is referred to as *perfect secrecy* [119] which requires the encryption key to be at least as long as the message.

- **Complexity-theoretic security**: An appropriate model of computation is defined which is used for security proofs. The adversary is assumed to have polynomially-bounded computational power in time or space. Asymptotic analysis and worst-case analysis are usually used in this context, which gives an upper bound on complexity of the problem and may affect the practical significance of the proofs. Finding a lower bound of a problem requires making a statement about all possible algorithms for solving the problem which may make it quite hard to achieve.

Complexity-theoretic security analysis may not have practical significance, but they give better understanding of the security.

- **Provable security**: A cryptographic scheme is *provable secure* if it can be proven that breaking the scheme is as difficult as solving a well-known hard problem. The proofs are usually done through a reduction to a hard problem. Provable security can be considered as a subclass of the computational security model.

- **Computational security**: A cryptographic scheme is computationally secure if the required computational efforts for breaking the scheme exceeds the hypothetical adversary's resources by an adequate margin. This class is sometimes called *practical security*.

- **Ad hoc security**: This approach consists of arguments which state any successful attack requires resources that exceed a potential adversary's resources. In this approach, primitives and protocols are examined against a list of standard attacks, e.g. those discussed in Section 2. Cryptographic primitives and protocols that survive such analysis are said to have *heuristic security* where claims of security remain questionable and unforseen attacks remain a threat [16].

## 3.1 Information-theoretic vs computational security

From the adversarial power limitations' perspective, one may classify all the aforementioned security models into two broad divisions that stand in contrast to each other: *information-theoretic security* and *computational security*. A cryptographic scheme is information-theoretically secure if an adversary with unlimited computational resources does not have enough information to succeed in its attack. In contrast, in computational security, an adversary has limited computational resources, but is allowed to eavesdrop and gain some information. The computational security is weaker than the information-theoretic security, but it is more realistic. Modern cryptographic schemes can be broken given enough time and computational resources, but the required time for breaking them should be more than the lifetime of the information that should be kept secret, even using all the available computational resources. The computational security relies on unproven assumptions, but the information-theoretic security does not require such assumptions. Any unconditional proof for computationally secure cryptographic schemes requires proving that $P \neq NP$, which is still an unsolved problem in computer science [120–122], but may be proved in the future. However, the information-theoretic security enforces some requirements which affects the practicability of a cryptographic scheme. As an example, a perfectly (or information-theoretically) secure encryption scheme requires that the encryption key should be as long as the combined length of all messages ever encrypted using this key. In the computational security approach, definitions of the security are mathematically weakened, but more *practical* cryptographic schemes are obtained.

The computational approach incorporates two relaxations with respect to the perfect security: to consider only efficient adversaries that run in a feasible amount of time; and to have adversaries that can have a very small probability of success. Such relaxations can be evaluated by either the *concrete* or *asymptotic* approaches. In the concrete approach, the running time and maximum success probability of an adversary are quantified; while in the asymptotic approach, they are viewed as functions of a *security parameter* which provides a relaxation from hardware specifications of an adversary [81].

## 3.2 Idealized models in computational security

In the computational security model, the *standard model* is the model of computation in which the adversary is only limited by the amount of time and computational resources. Proving the security of a certain cryptographic scheme is equivalent with proving a lower bound on the hardness of a certain computational problem [123]. Schemes which their security proofs are only based on complexity assumptions are said to be secure in the standard model. Proofs in the standard model are usually difficult to achieve. Then, in many proofs, cryptographic primitives are replaced by idealized versions.

The most popular instance of such idealization is the *random oracle model*. A random oracle is a random mathematical function which maps each unique query to a fixed random response from its output domain. Random oracles have been well-studied in computational complexity theory [124], and have been used in different cryptographic schemes [125–127], but their first explicit formalization for security proofs of cryptographic protocols was due to Bellare and Rogaway [128]. The random oracle model is an ideal system in which all parties (including the adversary) have oracle access to a truly random and publicly accessible function. In practice, random oracles are instantiated with a concrete cryptographic hash function. Random oracles are typically used as an ideal replacement for the hash functions in security proofs of schemes which involve hash functions and require strong assumptions for the hash functions. The adversary then must query the random oracle instead of calculation the hash function by himself. Many schemes involving hash functions need the random oracle model for security proofs, but this is not the case for any scheme. As an example, the Cramer-Shoup cryptosystem which deploys a universal one-way hash function family can be proved to be IND-CCA secure in the standard model under decisional Diffie-Hellman (DDH) assumption [129].

Using random oracles in security proofs is controversial, as no concrete hash function can achieve the functionality of a random oracle. In particular, there exist secure ideal encryption and signature schemes that are proven secure in the random oracle model, but insecure when the random oracle is replaced by any polynomial-time computable function [130]. However, security proofs in the random oracle model provide a common heuristic for practical security of a scheme, as the scheme is expected to remain secure if the random oracle is instantiated with a concrete hash function. Such proofs generally show that for breaking the protocol, an adversary requires impossible behavior from the oracle, or to solve some hard problems. It has been argued that the need for the random oracle assumption in a proof does not indicate a real-world security weakness in the corresponding protocol [131]. The leaky random oracle model [132] is a variant to the random oracle which allows adversaries to obtain contents of the hash list of input and output pairs arbitrarily.

The *ideal cipher model* is another idealized model of computation, similar to the random oracle model. It is has been used for different purposes from analysis of several constructions in symmetric cryptography [133–135] to studying generic related-key attacks [136]. In this model, all parties (including an adversary) are granted access to a publicly accessible random block cipher (or ideal cipher), and can make encryption and decryption queries to the ideal cipher. An ideal cipher is a random permutation oracle for modeling an idealized block cipher. It is possible to construct artificial schemes that are secure in the ideal cipher model, but insecure for any concrete block cipher [137]. Again, it provides good heuristics for the practical security. The ideal cipher model and the random oracle model can be considered equivalent [138]. Security in the random oracle model implies security in the ideal cipher model if one replaces a random oracle by a block cipher-based construction. For constructing an ideal cipher from a random oracle, a Feistel construction with 14 rounds may be used [139].

Another idealized model is the *generic group model* [140, 141] that was introduced for giving lower bounds on difficulty of the discrete logarithm and related problems [140], and has been used in security proofs of certain asymmetric encryption and signature schemes [142–144]. In this model, the adversary does not deal with efficient encodings such as multiplicative groups on finite fields or additive groups on elliptic curves, but is given access to a randomly-chosen encoding of a group. The adversary has then no information about the specific representation of the group being used. The generic group model has the same weakness as the random oracle model. There exist problems that are provably hard in the generic group model but easy to solve if the random encoding function is replaced with a specific encoding function [145]. The model can be extended to other algebraic structures such as rings which gives the *generic ring model* [146].

Other idealized models include the *common reference string (CRS) model* [147] which is a generalization to the *common random string model* in terms of adding the possibility of having strings with non-uniform distribution. The CRS model is widely used in cryptographic protocols, often for constructing non-interactive zero-knowledge (NIZK) proofs. It allows construction of cryptographic protocols that are provably impossible to realize in the standard model, for example the universally composable commitment [148]. The *multi-string model* [149] is an extension to the CRS model, mainly in terms of having a number of authorities, instead of having a trusted party for generating a random string. Those authorities will assist the protocol execution by providing random strings through a multi-party computation.

## 3.3 Formal security models

A formal security model must specify how an arbitrary probabilistic polynomial-time (PPT) adversary can interact with legitimate users of a cryptosystem, and what the adversary should achieve in order to break the cryptosystem. Formal security models can be generally categorized into *game-based* and *simulation-based* security models.

Games are programs, written in pseudocode or in some formalized programming language [134]. As mentioned in Section 2.2, security of cryptographic primitives and protocols can be formalized into game-based security models. Game-based security models have been introduced for different schemes including digital signatures [150], asymmetric encryption [58], symmetric encryption [82], and key exchange protocols [151, 152]. In game-based security models, a game is played between an adversary and a hypothetical *challenger*. A challenger is a probabilistic algorithm that generates all the keys used in the cryptosystem, and may respond to queries made by the adversary. The game terminates by an adversary's termination, and the adversary's queries and actions are then evaluated for meeting conditions for breaking the cryptosystem. The security of a cryptosystem is proven by showing that an adversary's probability of success is small. Game-based security models are easy to understand and deal with, but it is difficult to use them for complex protocols. Moreover, they make no claim about the security of a scheme when it is part of a larger system or environment, which is usually the case. Security of an isolated protocol instance cannot justify the security in a larger environment or in interaction with other protocol instances. Adversarially-coordinated interactions between different protocol instances can compromise the security of protocols that were shown to be secure when run in isolation [46, 153].

The *game hopping* [134, 154] is a technique used in this context as a tool for organizing a proof, though it is not applicable to all security proofs [154]. In a game hopping proof, an adversary running in a particular attack environment has an unknown probability of success. The attack environment is then altered slowly until the adversary's probability of success can be computed. For each alteration, there will be a change in the adversary's probability of success. A bounding is then deduced for the adversary's success probability in the original environment. In theory, games can change in any way. However, recognised types of game hop include bridging steps, transitions based on indistinguishability, transitions based on small failure events, and transitions based on large failure events [155].

Simulation-based security models stem from the simulation paradigm [156, 157]. They use the notion of *simulatability* between an *ideal* system that can never be broken, and a *real* system. Outputs from both systems should be simulatable. An arbitrary

PPT adversary interacts with each algorithm of the cryptosystem, and also with an arbitrary PPT *environment*. The environment represents all other entities that may have access to the algorithms of the cryptosystem. For security analysis, the adversary and the environment interact with both real and ideal cryptosystems, and their outputs are examined. Since the ideal cryptosystem cannot be broken, if the outputs of the environment and adversary in the real cryptosystem are roughly the same as those of the ideal cryptosystem, then the real cryptosystem should be secure. The security proof is accomplished by formulating the inability to distinguish between these two outputs. Examples of simulation-based models include Shoup's model for key exchange [158], Canetti's universally composable (UC) framework [159], and Pfitzmann and Waidner's composed systems [160]. A protocol is *universally composable* if it remains secure when composed concurrently with an unbounded number of instances of arbitrary protocols [153].

Simulation-based security models are stronger than game-based security models. They provide strong security guarantees, and consider the larger environment in which the cryptosystem will be deployed. Specifically, they can guarantee the *universally composable security* of a cryptographic protocol [159]. However, there are certain cryptographic functions that their security can never be proven by simulation-based security models [148]. Game-based and simulation-based approaches have their pros and cons, and neither provide a full range of applicability and security guarantees.

## 3.4 Security proofs in reality

Current methodologies used in provable security are controversial [161]. Most techniques used in security proofs do not help in protocol design. A small change in the protocol may require constructing a new proof. Security proofs include several pages of mathematical reasoning which makes them boring for many people. Another concern is the relationships between a security model, idealizations and assumptions in proofs, and the real world. A security proof depends on definitions and assumptions being used. If the threat model does not capture the adversaries' capabilities in the real world, or if the security guarantees do not match the requirements, or if the assumptions are incorrect, the security proofs are irrelevant and meaningless. Another concern in this regard is dependability of many proofs on idealized models mentioned in Section 3.2. The provable security does not necessarily imply the security in the real world. It just implies the security according to definitions and assumptions that the security model is built on.

Correctness of security proofs, and their ability in preventing basic errors are another concern. There are some protocols with formal proofs of security that could not even prevent basic errors, and the security proofs and claims turned out to be incorrect. The JP-MAKEP protocol due to Jakobsson and Pointcheval [162] was formally proved in the Bellar-Rogaway (BR) security model [163], but was later shown to be vulnerable to an impersonation attack [164]. Other examples of such protocols with formal proofs of security but vulnerability to basic attacks include Wong et al.'s WC-MAKEP protocol [164] and Bellare et al.'s encryption-based MT-authenticator [165] that are vulnerable to different attacks [166, 167]; Abdalla et al.'s 3PAKE protocol [168] which is vulnerable to a UKS attack [169] and an undetectable online dictionary attack [170]; and Abdalla et al.'s general construction [171] which is vulnerable to an undetectable online dictionary attack [170]. This might bring into question the value of security proofs in helping to avoid even basic errors, and shows that there are a few people who check the claimed proofs in detail [19].

## 4 Security models for cryptographic protocols

A non-exhaustive list of standard attacks on cryptographic protocols were reviewed in Section 2. Valid attacks on protocols are defined through security models, and security proofs provide reasoning about their provision, often by reduction to an assumed hard problem. Hard problems in most popular cryptographic protocols and public-key primitives are number-theoretic problems such as discrete logarithm problem and factorization that are infeasible by digital computers, but feasible with a quantum computer. Specifically, the Shor's algorithm [172] can perform integer factorization and discrete logarithms in polynomial-time on a quantum computer. *Quantum cryptography* offers some quantum key distribution schemes such as the BB84 protocol [173] which is provably secure [174, 175] or more recently, the Sasaki et al.'s scheme [176]. However, it is still far from popularity of quantum cryptosystems in the near future [177]. If large-scale quantum computing becomes technologically feasible, many cryptographic primitives and protocols will therefore be rendered completely insecure. This is a great motivation for the *post-quantum cryptography* [178] which tries to develop problems that are not feasible by quantum computers, and to develop efficient schemes based on hard problems that are already known to be intractable by a quantum computer. Current post-quantum solutions for public-key primitives and protocols are mainly based on lattices [179, 180], coding theory [181], hash functions [182], multivariate polynomials [183], and isogenies of elliptic curves [184]. A couple of quantum attacks on both classical and quantum cryptographic protocols have been proposed in the literature [185]. There are also proposals for post-quantum key exchange protocols [186].

Cryptographic protocols are prone to different kinds of design mistakes. Many protocols are proposed to fix problems in existing ones, but are later shown to have the same or other security flaws. A security flaw can be described by an attack scenario whereby some desirable or intended security services are shown to be failed. Reported attacks on new protocols improve our understanding of protocol design. There is no general rule for how to design a cryptographic protocol. However, there are some proposals for design principles in the literature [2, 187], mostly based on observations from common errors in published protocols. Those proposals do not guarantee that the outcome will be a secure protocol, but they can help designers to avoid repeating a number of published confusions and mistakes.

The design of secure protocols that provide certain desired functionalities or security services for different applications is a major part of modern cryptography. From another perspective, the design of any cryptographic scheme may be viewed as the design of a secure protocol for implementing a suitable functionality [157]. From the historical point of view, cryptographic protocols have had a great role in development of the public key cryptography. The first public key cryptosystem (RSA) [188] was introduced about one year after introduction of the Diffie-Hellman key exchange protocol [189]. Since then, many public-key schemes have been introduced whose security rest on different computational problems.

Cryptographic protocols can be designated for specific functionalities such as identification or key exchange protocols [19], specific requirements such as RFID protocols [190–192], specific applications such as secure SMS [193] or secure email [194], specific paradigms such as leakage-resilient protocols [87, 114], or they can be general protocols with many applications. In principle, many cryptographic protocols can be formulated as the *multi-party computation* (MPC). Secure MPC protocols are then sometimes referred to as the *general cryptographic protocols* [157]. As an example, in key exchange protocols, each principal has some secret values, and wants to securely compute a session key without compromising the secret values. In secure MPC, a set of $m$ parties, each having a secret value $x_i$, want to compute a joint function $f(x_1, ..., x_m)$, without revealing any information about $x_i$. Correctness of the computation and privacy of the inputs are two important objectives for the secure MPC. Secure two-party computation was introduced by Yao for semi-honest participants [195]. Since then, the MPC has been an active area of research, where research mostly focuses on designing general-purpose protocols, optimizing protocols for particular functions of interest, and implementing secure MPC [196, 197].

Notions of security have been defined for the MPC protocols both in computational and information-theoretic security [198, 199]. The simulation-based security models, introduced in Section 3.3, is actually the model used for security evaluation of general cryptographic protocols. As explained in Section 3.3, a real and an ideal model are defined. The ideal model for the MPC protocols consists of a trusted party which privately receives inputs from participants, and computes the function $f$ for them. In the real model, the participants compute $f$ without any trusted party. The protocol is deemed secure if the real setting emulates the ideal setting, that is whatever the adversary can obtain in the real setting can also be obtained in the ideal setting. Although security models developed for general protocols would be used for any cryptographic protocol, it is easier and more efficient to deal with security models that are specifically developed for a certain type of protocol. The most specialized and well-developed security models for analysis of cryptographic protocols in the computational setting are devoted to AKE and PAKE protocols that will be briefly reviewed in Section 4.1.

All the security models considered until now, stem from the cryptographic community. However, there are other methods for analysis of security protocols, known as the *symbolic models* which stem from the formal methods community, and are credited back to the seminal paper of Dolev and Yao [200], known also as the *Dolev-Yao model*. The computational models are originated from papers by Goldwasser and Micali [72] and Yao [201], and provide stronger security guarantees. We do not go through the symbolic models, but we will review differences between the computational and symbolic models in Section 4.2 for completeness of the discussion. There are efforts for bridging the gap between these two models, mostly through the computational-soundness theorems [202, 203].

## 4.1 AKE protocols

Authentication and key establishment are fundamental building blocks for providing different security services in secure communications. *Key establishment* is a process whereby a secret key is shared between two or more entities, for subsequent cryptographic use. Based on the number of entities involved in generating the session key, a key establishment protocol can be a *key transport*, a *key agreement*, or a *hybrid* protocol [16, 19]. In a key transport protocol, one principal generates the key, and the generated key is transferred to other principals. In a key agreement protocol, all the corresponding principals are involved in the key generation. In a hybrid protocol, more than one principal (but not all the principals) are involved in the key generation.

AKE protocols are the most well-studied type of security protocols. In 1976, Diffie and Hellman [189] introduced a key agreement protocol which is vulnerable to a MITM attack due to lack of authentication. Modern study of authentication and key establishment protocols is sometimes credited back to Needham and Schroeder's paper [204]. It can be formally proved that for an authenticated key establishment, we need some secure channels [205]. Excluding the case for a secure physical key establishment, we require that some secret keys are pre-shared between principals or with some trusted authorities, or principals should have certified public keys. AKE protocols should provide certain security attributes, and they should withstand well-known attacks introduced in Section 2.1. The *freshness* is an important attribute in key exchange protocols. The established key should be new, and not replayed from previous sessions. The freshness can be provided using timestamps, nonces or counters [28]. A key agreement protocol is expected to provide forward secrecy, known-key security, and joint key control [206], and to be resilient to the Denning-Sacco attack [17, 207]. Forward secrecy preserves the security of session keys after disclosure of keying material, used in the protocol to negotiate session keys. Known-key security means that each run of a protocol should produce a unique session key. Compromise of a session key should not compromise other session keys. Joint key control ensures that all the intended participants are involved in generating the session key, and ensures that no entity is able to enforce the session key to fall into a pre-determined interval [208]. Resilience to Denning-Sacco attack prevents an adversary to recover or guess the secret parameters used in the protocol upon disclosure of a session key.

### 4.1.1 Security models for AKE protocols

Although AKE protocols can be considered as special cases of the MPC, many security models have been specifically developed for AKE protocols in the computational setting. The simulation-based security models have been discussed in Section 3.3. An example is the Shoup model for key exchange [158]. The first game-based security model for AKE protocols was the Bellare-Rogaway (BR93) model [163] which covered mutual authentication and key exchange from pre-shared symmetric keys. Later, Bellare and Rogaway introduced the BR95 model [151], which is an extension to the BR93 model and covers server-based key exchange. In the BR models, the security of a KE protocol is defined in terms of indistinguishability of established session keys from random values through a game played with a PPT adversary. The adversary has access to any public data and controls all the communications by interacting with a set of oracles, each of which represents an instance of a principal in a specific protocol run. The adversary interacts with the principals via queries which are mainly `Send`, `Reveal`, `Corrupt`, and `Test`. `Send` allows the adversary to make the principals run the protocol. `Reveal` models the adversary's ability to find old session keys. `Corrupt` models insider attacks by the adversary, returns the oracle's internal state, and sets the long-term key of a principal to a value chosen by the adversary. The adversary can then control the behaviour of the corrupted principal by `Send` queries. Success of the adversary is measured in terms of its advantage in distinguishing the session key from a random unrelated value after running the `Test` query. The BR models were a milestone, and led to other security models. Blake-Wilson and Menezes [209, 210] extended the BR models to cover public keys and key agreement. Bellare, Pointcheval, and Rogaway [211] extended and modified the BR95 model, and introduced the BPR model for password-based protocols. The BPR model will be briefly reviewed in Section 4.1.2.1.

#### 4.1.1.1 CK01 model

Bellare, Canetti and Krawczyk [165] introduced a general framework and a modular approach for design and analysis of authentication and key exchange protocols, which is sometimes referred to as the BCK98 model. Subsequently, Canetti and Krawczyk introduced the CK01 model [212] which fixed up problems with the BCK98 model, and provided a prototype of a modern security model by combining the BR and BCK98 models. Basic ideas in the CK01 model are similar to those of the BR models, but the CK01 model allows session states to be revealed which captures more security attributes. Following the BCK98 model, two adversarial models are defined in the CK01 model, namely the unauthenticated-links adversarial model (UM) and the authenticated-links adversarial model (AM). In the UM model, the adversary is active and has full control over the communication links; while in the AM model, the adversary cannot fabricate messages and can only deliver messages truly generated by the parties without any change or addition. AM is an ideal setting, and UM is the real setting. Then, we have the notion of *emulation* to capture the equivalence of functionality between protocols in the UM and AM models. Using an appropriate authenticator, a protocol that is proven to be secure in the AM model can be transformed to a provably secure protocol in the UM model [165, 212].

A party may be engaged in multiple instances of the protocol, each called a *session*. Any party $P_i$ starts a protocol run when it receives an input of the form $(P_i, P_j, s, role)$ in which $P_i$ is the owner of the session, $P_j$ is the peer of the session, $s$ is the session identifier, and *role* can be either *initiator* or *responder*. Partnership in the CK01 model is defined through the notion of *matching sessions*. Two sessions are said to be *matching* if they have the same session identifiers. However, there is no concrete definition of session identifiers in the CK01 model. Session identifiers are chosen by a higher layer protocol which calls the protocol. The calling protocol should make sure that a party should never participate in two distinct sessions with the same session identifiers. The adversary controls all communications between parties as well as activation and expiration of sessions. The adversary is allowed to reveal session-specific secret information of a session (via `Session-state reveal` query), to reveal the session key of a session (via `Session-key query`), and to take full control of any party (via `Party corruption` query). The `Session-state reveal` query can be asked of an incomplete session, and returns the internal state. It includes practical scenarios where an adversary can gain access to the ephemeral data. The `Session-key query` reveals the session key of any completed session to the adversary. The `Party corruption` can be called at any point whereby the adversary learns all the internal memory of that party including long-term secrets (such as private keys or shared master keys used across different sessions) and session-specific information contained in the party's memory (such as internal state of incomplete sessions and session keys of completed sessions).

An important element in the CK01 model is the notion of *session expiration* which helps in formalization of the perfect forward secrecy (PFS). A session is expired if the session key agreed by the session has been erased from the session owner's memory. By the `expire-session` query, an adversary directs a completed session to delete its session key. For a session to be locally exposed, the attack against the session must happen before the session expiration. The CK01 model allows a session to be expired at one party without necessarily expiring the matching session. A session is said to be *locally exposed* if the adversary issued a `Session-state reveal` or `Session-key reveal` query to the session, or if the adversary issued a `Party corruption` query to the session owner before the session is expired (before asking a `expire-session` query for the session). A session is called *exposed* if a session or its matching session is locally exposed. A session is called unexposed (or fresh) if neither the session nor its matching session are locally exposed.

The security of a protocol in the CK01 model is defined based on inability of a PPT adversary in winning a distinguishing game. The adversary's goal is to distinguish a session key from a random value. In the security game, the adversary can send messages to any instance and observe the response. The adversary can also ask `Session-state reveal`, `Session-key reveal`, `Party corruption`, and `expire-session` queries according to specifications of each query. At some point, the adversary selects an unexposed (fresh) session, and issues a `Test` query to that session. The adversary can continue to issue other queries as long as the test session remains unexposed. In response to the `Test` query, the adversary is given with equal probability either the session key held by the test session or a random key. The adversary is said to win the distinguishing game (which means breaking the protocol) if it can guess with non-negligible success probability greater than $1/2$ whether the key is random or not. A key agreement protocol is called session key (sk-) secure in the UM if (i) The protocol satisfies the property that if two uncorrupted parties complete matching sessions then they both output the same key; and (ii) the advantage of an efficient adversary in winning the distinguishing game is negligible. If the above properties are satisfied for all KE-adversaries in the AM, the protocol is said to be *SK-secure in the AM* [212]. In the above definition, the condition on freshness which allows the `Party corruption` after the `expire-session` preserves the forward secrecy. However, the CK model includes a variant without PFS for which the notion of *SK-security without PFS* is introduced that simply disallows the adversary to corrupt a partner to the test session.

SK-secure protocols in the CK01 model are resilient to the UKS attacks. The proof is straightforward by showing a contradiction. Suppose that a SK-secure protocol is vulnerable to a UKS attack where an entity $P_i$ is convinced to share a key with $P_j$ while the key is actually shared with another entity $P_{j'}$. Then, there will be two sessions $(P_i, P_j, s, role)$ and $(P_{j'}, P_i, s, role')$ that are non-matching because of different identifiers for the corresponding participants, but have the same session key. The adversary can then select one of those sessions as the test session, and reveal the session key of the other session. This actually reveals the session key of the test session. As those two sessions are non-matching, this is allowed and contradicts with the original assumption that the protocol is SK-secure.

The CK01 model has been used for security analysis of many AKE protocols. However, it has received some criticism. As mentioned earlier, the CK01 model does not provide a concrete definition of the session identifiers. Furthermore, definition of the session state is up to protocol designers which can cause ambiguities in security proofs of protocols and their implementations. Specifically, the `Session-state Reveal` query reveals the internal state of the Turing machine executing the protocol to the adversary, but this internal state is not defined in the CK01 model. This implies that the proofs are valid only for execution models corresponding to the specific definition of the internal state in the proofs. As definition of the internal state is not usually explicitly stated as a parameter for correctness of proofs, this puts restrictions on implementations of the protocol. Any implementation in which the local state (as revealed to the adversary) contains more information than the corresponding definition in the proofs, falls outside of the scope of the proof [213]. Another issue is that the CK01 model does not capture some important attacks such as the KCI attack and its variants. The `Session-State Reveal` query in the CK01 model allows an adversary $\mathcal{M}$ to have the state-specific information of the parties, but any session which has had a `Session-state Reveal` query is not fresh and cannot be the test session. Then, the CK01 model does not capture the KCI attacks which allow an adversary to reveal some secret parameters for the test session. Resilience to the KCI attack is an important security attribute for AKE protocols. If the private key of an entity $\mathcal{A}$ is compromised, an adversary $\mathcal{M}$ can impersonate $\mathcal{A}$ in one-factor authentication protocols. However, such compromise should not enable $\mathcal{M}$ to impersonate other honest entities in communication with $\mathcal{A}$ [214].

Generally, the computational models differ based on definition of partnering and freshness, what the adversary is allowed to obtain, and when the adversary is allowed to obtain things (ask queries). A comparison between the BR93, BR95, BPR and CK01 models is provided in [169].

The BR and CK01 models can be categorized into the *pre-specified peer model* wherein it is assumed that a party knows the identifier of its intended communicating peer when it commences a run of the protocol. Canetti and Krawczyk [215] also introduced a variant of the CK01 model in the *post-specified peer model* where a party only knows a destination address of its communicating peer, and learns the peer's identifier during the execution of the protocol. In practice, this happens where the peer's identifier is simply unavailable at the outset or if one party wishes to conceal its identity from adversaries. An example of protocols with option of identity concealment is the IKE protocol, used in the IPSec [215, 216]. It has been shown that security in the Canetti-Krawczyk's pre-specified peer model does not guarantee the security in the Canetti-Krawczyk's post-specified peer model, and vice versa [217]. Canetti and Krawczyk [218] also tried to bridge between seemingly limited indistinguishability-based definitions such as the SK-security and more powerful, simulation-based definitions, such as the UC security, where general composition theorems can be proven.

#### 4.1.1.2 HMQV model

As mentioned earlier, the CK01 security model does not capture the KCI attacks. This has been considered in a security model which was introduced for security analysis of the HMQV protocol [207]. After introduction of the MQV protocols [206] that are a family of efficient and widely standardized AKE protocols but without formal proofs of security, Krawczyk [207] reported several security problems in the MQV protocols, and proposed the HMQV (hashed variant of the MQV) protocols. Despite the MQV's security problems reported in [207], Kunz-Jacques and Pointcheval [219] presented security proofs for the MQV protocol in a very restricted security model, and claimed that proper key derivation is enough to overcome security weaknesses of the plain MQV protocol. Design and security proofs of the HMQV protocol were based on a new form of challenge-response signatures which was derived from the Schnorr identification scheme [126, 220]. The HMQV protocol was claimed to provide all the MQV's security goals in the random oracle model under the computational Diffie-Hellman (CDH) assumption [207], by showing the security of the exponential challenge-response (XCR) signature in the random oracle model and invoking the forking lemma [221] and then using it to imply the security of the HMQV protocol. For achieving better performance, public-key validations that were mandated in the MQV protocols, were omitted from the HMQV protocols. This makes the HMQV protocol vulnerable to the small-subgroup attacks which allows an attacker to recover a victim's static private key [50, 51]. Although the HMQV security model somehow ignored the modular approach in the CK01 security model, but it considered some other notions such as resilience to the KCI attack and weak perfect forward secrecy (wPFS).

#### 4.1.1.3 eCK model

The eCK (extended-CK) model was introduced by LaMacchia et al. [152], and is an extension to the CK01 model. It tackles some weaknesses in the BR and CK01 models. Specifically, the adversary can obtain ephemeral secrets which belong to the test session. The adversary can obtain the long-term key of the test session and of its partner even before the session is completed. The eCK model allows various combinations for leakage of long-term and ephemeral secret keys, but not both leakages happening at the same entity. Let $sk_\mathcal{A}$ and $sk_\mathcal{B}$ denote long-term secret keys of Alice ($\mathcal{A}$) and Bob ($\mathcal{B}$), respectively. Let $esk_\mathcal{A}$ and $esk_\mathcal{B}$ be ephemeral secret keys of $\mathcal{A}$ and $\mathcal{B}$, respectively. From different combinations for revealing those secrets, the adversary is allowed to obtain one of the following combinations: ($esk_\mathcal{A}$, $esk_\mathcal{B}$) or ($sk_\mathcal{A}$, $sk_\mathcal{B}$) or ($esk_\mathcal{A}$, $sk_\mathcal{B}$) or ($sk_\mathcal{A}$, $esk_\mathcal{B}$). However, the adversary is not allowed to obtain either both $sk_\mathcal{A}$ and $esk_\mathcal{A}$, or both $sk_\mathcal{B}$ and $esk_\mathcal{B}$.

As in the CK01 model, partnership in the eCK model is defined through session identifiers, and we have the notion of *matching sessions*. However, sessions are only partners if they agree on which one of them takes the initiator and which takes the responder role. In the eCK model, the session identifier $sid$ is defined as concatenation of the parties' identities and the messages they exchange in the session. Specifically, $sid = (role, ID, ID^*, comm_1, ..., comm_n)$ where $ID \in \{0,1\}^*$ is the identity of the party executing the session, $role \in \{\mathcal{O}, \mathcal{P}\}$ is its role as either the owner or the peer, $ID^*$ is the identity of the other party, and $comm_i \in \{0,1\}^*$ is the i-th communication sent by the parties. The matching session to a session (by the owner with the peer) is defined as the corresponding session which is supposed to be executed by the peer with the owner. For example, if $sid = (\mathcal{O}, \mathcal{A}, \mathcal{B}, comm_\mathcal{A}, comm_\mathcal{B})$ is a session executed by $\mathcal{A}$, then the matching session is executed by $\mathcal{B}$ and has session identifier $sid^* = (\mathcal{P}, \mathcal{B}, \mathcal{A}, comm_\mathcal{A}, comm_\mathcal{B})$. The matching session might not exist if the communications were modified by the adversary [152].

In order to enable the adversary to reveal ephemeral secrets, the new `Ephemeral Key Reveal` query is defined in the eCK model. The `Session-state Reveal` query of the CK model does not exist in the eCK model. In the eCK security model, the adversary can make any sequence of the following queries during the indistinguishability game [152]:

- `Send`$(\mathcal{A}, \mathcal{B}, comm)$: By this query, the adversary orders $\mathcal{A}$ to start a session with $\mathcal{B}$, and provides communications from $\mathcal{B}$ to $\mathcal{A}$. It sends a message *comm* to $\mathcal{A}$ on behalf of $\mathcal{B}$, and returns $\mathcal{A}$'s response to this message.

- `Long-Term Key Reveal`$(\mathcal{A})$: This query reveals the long-term key of $\mathcal{A}$.

- `Ephemeral Key Reveal`$(sid)$: This query reveals an ephemeral key of a (possibly incomplete) session $sid$.

- `Reveal`$(sid)$: This query reveals a session key of a completed session $sid$.

Eventually at any time in the experiment, the adversary can make only one query `Test`$(sid)$ for a completed session $sid$. The adversary wins the experiment if the selected test session is *clean* (fresh), and if it could guess the challenge correctly.

Let $sid$ be an AKE session completed by a party $\mathcal{A}$ with another party $\mathcal{B}$. Let $sid^*$ denote the matching session to $sid$. An AKE session $sid$ is not *clean* if any of the following conditions holds [152]:

- $\mathcal{A}$ or $\mathcal{B}$ is an adversary-controlled party. $\mathcal{A}$ is called adversary-controlled if both $sk_\mathcal{A}$ and $esk_\mathcal{A}$ are revealed.

- The adversary reveals the session key of $sid$ or $sid^*$ (if $sid^*$ exists).

- If $sid^*$ exists, the adversary reveals either both $sk_\mathcal{A}$ and $esk_\mathcal{A}$, or both $sk_\mathcal{B}$ and $esk_\mathcal{B}$.

- If $sid^*$ does not exist, the adversary reveals either $sk_\mathcal{B}$, or both $sk_\mathcal{A}$ and $esk_\mathcal{A}$.

The eCK model was intended to be stronger than the CK01 model. Here are some practical attack scenarios that are not considered in the CK01 model, but they are captured by the eCK model [152]:

- KCI attack: An adversary reveals a long-term secret key of a party, and impersonates others to this party.

- Ephemeral KCI attack: An adversary reveals the ephemeral secret key of a party, and impersonates others to this party.

- Two honest parties execute matching sessions. Adversary reveals the ephemeral secret keys of both parties, and tries to learn the session key.

- Two honest parties execute matching sessions. Adversary reveals the ephemeral secret key of one party and the long-term secret key of the other party, and tries to learn the session key.

- Two honest parties execute matching sessions. Adversary reveals the long-term keys of both parties prior to the execution of the session, and tries to learn the session key.

However, the CK01 and eCK models seem *incomparable* [213, 222, 223]. Although in some literature [152, 224, 225], the `EphemeralKeyReveal` query of the eCK model has been deemed to be at least as strong as the `Session-StateReveal` query of the CK01 model, it has been argued in other literature [213] that the `Session-StateReveal` query is stronger than the `EphemeralKeyReveal` query. Specifically, the argument is supported by showing that the NAXOS protocol [152], which was proven to be secure in the eCK model, is vulnerable to some attacks in a variant of the eCK model (called eCK' model) which is obtained by replacing the `EphemeralKeyReveal` query of the eCK model with the `Session-StateReveal` query of the CK01 model [213]. The NAXOS protocol uses an idea known as the NAXOS trick, where the ephemeral secret key is combined with the long-term key. This prevents the adversary from performing a successful attack as the adversary will need knowledge of both ephemeral and long-term keys that is not allowed in the eCK model. The trick was bypassed using the `Session-StateReveal` query in the eCK' model where the session state includes all the parameters except the long-term key. A comparison between CK01, HMQV, and eCK models is provided in [223] which concludes these models are incomparable, i.e. security in each of these three models does not imply security in the other two models.

The notion of matching sessions in definition of partnership in AKE security models can be used for constructing a *key-replication* attack which invalidates the semantic security of a protocol in the corresponding models. In a key-replication attack [207, 226], an attacker establishes two different non-matching sessions so that they output the same session key. The adversary then selects one of the two non-matching sessions as the test session which will remain unexposed. The other non-matching session is then queried by the adversary to reveal its session key. As the queried session has the same session key as of the test session, the adversary could indeed obtain the session key of the test session which is still unexposed. If this attack is successful, the adversary can simply win the indistinguishability game, and it invalidates the semantic security of the protocol.

#### 4.1.1.4 eCK-PFS model

One of the controversial aspects of the HMQV and eCK security models is their relation to the PFS. Forward secrecy is an important security attribute in AKE protocols, and it is somehow addressed by many security models. If long-term keys of entities are compromised, it should not affect the security of previous session keys established before the compromise [227]. A protocol provides forward secrecy if the adversary cannot distinguish the session key from a random string even given the long-term keys of both parties after the session is completed. Forward secrecy and weak forward secrecy were formalized in the BPR model. It was concluded in the BPR model [211] that forward secrecy in the strong-corruption model is not achievable by two-flow protocols. Later, during the HMQV model, Krawczyk [207] introduced an active attack that can defeat the PFS for a generic two-message key exchange protocol, which is authenticated via public keys and no secure shared state is perviously established between the entities. The HMQV model [207] then considered the notion of weak-PFS (wPFS) in which the adversary is not allowed to have an active part in the session under attack. wPFS allows an adversary to reveal static private keys of the involved entities, but the adversary is only allowed to accomplish a passive attack. The Krawczyk's generic PFS attack [207] led to a belief in the security community that no two-pass AKE protocol can achieve the PFS, and the best they can achieve is the wPFS [152, 207, 228, 229]. Such belief has been argued to be incorrect in [230] by introduction of eCK$^w$ and eCK-PFS security models where the eCK$^w$ model provides a slightly stronger variant of wPFS, and the eCK-PFS model is strictly stronger than the eCK$^w$. Then, a security-strengthening protocol transformation (or compiler) *SIG* can transform a generic two-message DH type protocol from eCK$^w$ model to a two-message protocol which is secure in the eCK-PFS model. The approach includes relaxing the notion of the matching session, and introducing the notion of the *origin-session* [230].

#### 4.1.1.5 Further models and protocols

A couple of protocols have been proposed as improvements to the MQV and HMQV protocols, and several security models have been introduced for those protocols. Examples include CMQV protocol in the eCK model [224], UP protocol [222] in Menezes-Ustaoglu (eCK+) model [217], SMQV protocol in seCK model [229], FHMQV protocol in the FHMQV model [231] (with arguments on incomparability to the HMQV model and flaws in security proofs [232]), and UP+ protocol in vCK model [233].

Most efficient protocols like the HMQV protocol rely on the random oracle model for their proofs. As mentioned in Section 3.2, it is good to avoid idealized models such as the random oracle model. Subsequently, many efforts have been done to construct protocols in the standard model, typically using kind of pseudo-random function (PRF) [234–236]. Other trends include security models and protocols for capturing bad randomness [237], DoS-resilient protocols such as DoS-CMQV [238] and DoS-BPV-JFK [239], leakage-resilient eCK-secure protocols [87, 240, 241], and post-quantum key exchange protocols [186].

Most of protocols that involve public keys require a public key infrastructure (PKI) for certifying public keys, although there are many proposals for identity-based [242] and certificateless [243] protocols that do not require a PKI. In a PKI, a certification authority (CA) takes care of issuing and managing certificates [244, 245]. Many PKI-based protocols assume that a CA will perform proof-of-possession (PoP) to ensure that an entity knows the corresponding private key of the claimed public key. This is a way to thwart a public key substitution UKS attack [26]. However, trusting a CA to perform such checking is not so realistic [246]. The eCK security model stressed that an adversary is allowed to register arbitrary public keys for adversary-controlled parties without any checks such as PoP done by the CA [152]. However, this has not been explicitly modeled through queries. Subsequently, the ASICS model [246] has been introduced which concerns the certification procedure in PKI-based protocols, and provides an adversary queries for registration of valid or invalid public keys. It allows entities to have several public keys or use the same public key with different identifiers.

Most key exchange protocols are designed for two participants. However, in some applications such as video conferencing and secure group communication, we have arbitrary number of parties. In a group AKE protocol (GAKE), multiple parties accomplish an AKE. Several protocols [19] and security models [247, 248] have been proposed in the literature for the GAKE. The ability of capturing the insider attacks [248] has been considered in most recent security models for GAKE.

The authentication is usually based on knowledge factors (or what a user knows, e.g. a password), possession factors (or what a user has, e.g. a security token), and inherence factors (or what a user is, e.g. biometric data). Many AKE protocols use one of these factors for authentication and key establishment. However, multi-factor AKE protocols [249] use multiple factors for authentication and key establishment, and offer very strong protection for secure communication in highly sensitive applications.

### 4.1.2 PAKE protocols

Password-based AKE (PAKE) protocols enable two or more entities to authenticate each other, and share a cryptographic key based on a pre-shared human-memorable password. Due to low-entropy of passwords, such protocols are prone to online password guessing attacks, which can be prevented by limiting the number of failed trials at the server side. The goal in PAKE protocols is to ensure that the only feasible attack is a detectable online password guessing attack. PAKE protocols should be resilient to offline and undetectable online password guessing attacks [17, 45].

The encrypted key exchange (EKE) protocol was the first PAKE protocol, proposed by Bellovin and Merritt in 1992 [250]. In the EKE protocol, two users execute an encrypted version of the Diffie-Hellman key exchange protocol, where each flow is encrypted using the pre-shared password as the symmetric key. Since introduction of the EKE protocol, many PAKE protocols have been proposed [19], but many of them have security problems.

Some PAKE protocols are strictly password-only protocols, while others may require the authentication server to have a pair of private and public keys [69]. Based on the number of participants, a PAKE protocol can be categorized into two-party, three-party, four-party, or multi-party setting. In the two-party setting, the participants are two entities, typically a client and a server that have shared a password. In gateway-oriented PAKE protocols, an authentication server is seen as two entities, a gateway (which is directly in contact with the client and can be dishonest) and a back-end server (which verifies the identity of the client). In the three-party setting (C2C-PAKE), there are typically two clients that have shared passwords with a trusted server, and the goal is to have an authenticated key establishment between clients. In the four-party setting (cross-realm C2C-PAKE), two clients have shared passwords with different servers, and they want to have an authenticated key establishment. In the multi-party setting, password-based authenticated key exchange is accomplished between a group of participants.

Two-party PAKE protocols are a well-studied type of protocols, and many protocols and security models have been proposed in this setting. The first game-based security model for PAKE protocols was proposed by Bellare, Pointcheval, and Rogaway (BPR) [211], which is an extension to the BR95 security model [151]. The first simulation-based security model for PAKE protocols was proposed by Boyko, MacKenzie, and Patel (BMP) [251], which is an extension to Shoup's simulation-based security model [158]. The BPR model will be reviewed at the end of this section.

The first provably secure PAKE protocols in the standard model were the Katz-Ostrovsky-Yung (KOY) protocol [252] which is based on the decisional Diffie-Hellman assumption, but needs a common reference string; and the Goldreich-Lindell protocol [253] which is based on the general assumptions without any trusted setup assumption. Most PAKE protocols have security proofs in idealized models or assume the existence of a trusted common reference string. Protocols in the standard model without any trusted setup assumption [253, 254] are usually not efficient enough in practice.

Most of existing PAKE protocols have proofs either in the BPR model or in the BMP model. Although these models provide a level of security, they have limitations on the password distribution. Then, Canetti et al. [255] proposed an ideal functionality for PAKE protocols in the universally composable (UC) framework [159] where the environment emulates any distribution, mistypes of passwords and related passwords. However, it still fails to capture some leakage of information that can happen in reality.

The first formal security model for C2C-PAKE protocols was introduced by Abdalla et al. (AFP model) [171]. It is based on the BPR security model, but is stronger than the BPR model. The BPR model is based on the *find-then-guess* scenario where an adversary can ask only *one* Test query. An adversary's advantage in distinguishing between a random and real session key should not be significantly greater than $\mathcal{O}(q_s)/N$, in which $q_s$ denotes the number of active attacks, and $N$ is the size of the dictionary. However, the AFP model is based on the *Real-or-Random* scenario which allows an adversary to ask many Test queries, but the advantage should remain in $\mathcal{O}(q_s)/N$. Abdalla et al. [171] also provided a generic construction to transform any secure 2-party PAKE protocol into a secure 3-party PAKE protocol. To reduce the complexity of the generic construction, Abdalla and Pointcheval [168] proposed the 3PAKE protocol with provable security in the random oracle model, based on the AFP security model. However, their protocol was shown to be vulnerable to a UKS attack [169]. Wang and Hu [170] showed that both Abdalla et al.'s general construction [171] and the 3PAKE protocol [168] are vulnerable to undetectable online dictionary attacks [170], and proposed an enhancement to the AFP model for the treatment of undetectable attacks as well as a new generic construction for 3-party PAKE protocols. A couple of other C2C-PAKE protocols with provable security were shown to be vulnerable mostly to the UKS or undetectable online dictionary attacks [256]. Several other C2C-PAKE protocols and security models have been proposed in the literature [257].

The first gateway-oriented PAKE protocol was introduced by Abdalla et al. [258]. Abdalla et al. proposed two gateway-oriented PAKE protocols with provable security [258, 259], but both protocols are shown to be vulnerable to undetectable online password guessing attacks [260, 261]. Then, new security models and protocols have been introduced for capturing and thwarting the attack [261, 262].

Cross-realm C2C-PAKE allow clients to be in the realm of different servers. Several security models [263] and generic constructions [264] have been proposed for cross-realm C2C-PAKE protocols, and some cross-realm C2C-PAKE protocols with provable security have been shown to be flawed [265, 266].

In the multi-party setting, several group PAKE protocols have been proposed in the literature [267], many of them are based on the Burmester and Desmedt method [268]. They allow a group of participants to have an authenticated key establishment.

#### 4.1.2.1 BPR model

As mentioned earlier, the BPR model [211] was introduced as a variant of the BR95 model [151] for PAKE protocols. It aimed to deal with password guessing, forward secrecy, server compromise, and loss of session keys. In the BPR model, each principal $U$ is either a *client* or a *server*, and is named by a string of fixed length. Each principal $U \in ID$ where $ID$ denotes the union of finite and disjoint sets *Client* and *Server*. Each principal $A \in Client$ holds some password, $pw_A$. Each server $B \in Server$ holds a vector $pw_B = <pw_B[A]>_{A \in Client}$ which contains an entry per client. $pw_A$ and $pw_B$ are called long-lived keys (LL-keys). In the symmetric model, $pw_B[A] = pw_A$. In the asymmetric model, $pw_B[A] = pw_A$ and it is hard to compute $pw_A$ given $A$, $B$, and $pw_B[A]$. LL-keys of principals are determined by running a LL-key generator [211].

The instance $i$ of principal $U \in ID$ is called an *oracle*, and is denoted by $\Pi_U^i$. Each oracle might be embodied as a process, running on some machine controlled by the corresponding principal. An oracle may *accept* at any time, but it can accept at most once. When an oracle accepts, it holds a session key $sk$, a session identifier $sid$, and a partner identifier $pid$. The $SID$ is an identifier which uniquely names the corresponding session. If the protocol specification does not determine $SID$, it can be defined as concatenation of all protocol flows, i.e. messages exchanged during the protocol run. The $PID$ identifies the peer principal with which the instance believes it has exchanged a key. The session key is secret, but $SID$ and $PID$ are not secret. An instance which has accepted may still send out messages, until it terminates. An adversary has an oracle access to a function $h$ which is selected randomly from some probability space $\Omega$. The choice of $\Omega$ determines if the proofs are either in the standard model, random oracle model, or ideal cipher model. Difference between these models was reviewed in Section 3.2. The adversary has an endless supply of oracles, and can make the following queries to any instance of principals [211]:

- `Send`$(U, i, comm)$: This query sends message $comm$ to oracle $\Pi_U^i$. The oracle computes according to the protocol, updates the state, and sends the result back to the adversary. If this query causes $\Pi_U^i$ to accept or terminate, this fact will be made visible to the adversary. To initiate a session with client $A$ trying to enter into an exchange with server $B$, the adversary issues `Send`$(A, i, B)$ to an unused instance of $A$. This query allows an adversary to be active and fabricate any message.

- `Execute`$(A, i, B, j)$: This query performs an honest execution of the protocol between unused oracles $\Pi_A^i$ and $\Pi_B^i$, and outputs the transcript of the execution. This query models a passive attack, and is used for modeling a dictionary attack and differentiating between active and passive attacks.

- `Reveal`$(U, i)$: This query outputs $sk$ if oracle $\Pi_U^i$ has accepted and holds the session key $sk$. It can model the Denning-Sacco attack.

- `Corrupt(U, pw)`: This query can be used in different scenarios: (i) In the weak-corruption model, it returns the LL-key $pw_U$ to the adversary; (ii) In the strong-corruption model, it returns $pw_U$ and also the current state of all instances of $U$ to the adversary; (iii) When directed against a client $U$, it can be used to replace the value of $pw_B[U]$ which is used by the server. The strong-corruption model corresponds to completely compromising $U$'s machine, while the weak-corruption model corresponds to acquiring the password $pw_U$.

- `Test(U, i)`: If oracle $\Pi_U^i$ has accepted and holds the session key $sk$, a coin $b$ is flipped. If $b = 0$, then $sk$ is returned to the adversary. Otherwise, a random string from the space of the session key is returned. In the BPR model, the adversary can ask this query just once.

**Partnership in the BPR model**: Let two oracles $\Pi_U^i$ and $\Pi_{U'}^{i'}$ both accept, holding $(sk, sid, pid)$ and $(sk', sid', pid')$, respectively. They are called *partners* if all the following conditions holds [211]:

- $sk = sk'$, and $sid = sid'$, and $pid = U'$, and $pid' = U$,

- $[U \in Client$ and $U' \in Server]$, or $[U \in Server$ and $U' \in Client]$,

- No oracle besides $\Pi_U^i$ and $\Pi_{U'}^{i'}$ accepts with a PID of $pid$.

**Freshness in the BPR model**: Two notions of freshness are defined in the BPR model [211]:

- Basic notion of freshness (with no requirement for forward secrecy): An oracle $\Pi_U^i$ is *fresh* if all the following do not happen at any time:

  - `Reveal(U, i)`,
  - `Reveal(U', i')` where $\Pi_{U'}^{i'}$ is the partner to the oracle,
  - `Corrupt(U, pw)`.

- A notion of freshness with forward secrecy: An instance $\Pi_U^i$ is *fs-fresh* if all the following do not happen at any time:

  - `Reveal(U, i)`,
  - `Reveal(U', i')` where $\Pi_{U'}^{i'}$ is the partner to the oracle,
  - [`Corrupt(U', pw)` was made before the `Test(U, i)` query] and [`Send(U, i, comm)` was made for some string $comm$].

An oracle that has accepted can have at most one partner. The *AKE security* (with no requirement for the forward secrecy) is defined through a game wherein an adversary asks one `Test(U, i)` query for a terminated and *fresh* oracle $\Pi_U^i$, and outputs a bit $b'$.

For capturing forward secrecy, $\Pi_U^i$ must be *fs-fresh*. If $b' = b$ where $b$ is the output of the `Test` query, the adversary *wins*. Like the BR security model, the advantage of the adversary in attacking is defined as twice the probability that she wins minus one.

The BPR model provides guarantees for the authentication. An adversary violates client-to-server authentication if some server oracle terminates without having a partner oracle. An adversary violates server-to-client authentication if some client oracle terminates without having a partner oracle. An adversary violates mutual authentication if some oracle terminates without having a partner oracle. However, the BPR model is shown to be the weakest security model among BR93, BR95, and CK01 security models. It does not capture some attacks including the UKS attack [169].

## 4.2 Formal verification of Security Protocols

Formal verification of software correctness forms an important part of practical and theoretical computer science. Since security protocols can be viewed as short programs or algorithms, one may adapt software correctness methods and tools for verification of security protocols. However, complexity considerations, importance of security protocols, and the fact that the security problem in presence of an adversary cannot be detected by functional software testing indicate that there is a need for specialized tools.

Verification of a security protocol means verifying that the protocol is correct and performs according to its security objectives. Verification may indicate examples of failures or flaws in the analyzed protocol. A famous example is the Needham-Schroeder protocol [204] on which a man-in-the-middle attack was found using the FDR model checker, seventeen years after publication of the protocol [269], although correctness of the protocol was formerly verified using the BAN logic [270]. Concurrent executions of protocols where an entity may have different roles in different executions (for example as an initiator or a responder), and multi-protocol attacks mentioned in Section 2.1 make the analysis very complicated which cannot be captured by a heuristic verification. The verification problem is undecidable in its most general form [271, 272]. For unbounded message size in presence of an active adversary or unbounded number of sessions, the state space to explore is infinite and the problem is undecidable [273]. However, the secrecy preservation is NP-complete for a bounded number of protocol sessions [274] with respect to the Dolev-Yao model [200], and decidable for an unbounded number of sessions under some additional restrictions [203].

Formal methods, as defined by Meadows [275], is a combination of a mathematical or logical model of a system and its requirements, together with an effective procedure to determine whether a proof that a system satisfies its requirements is correct. Formal verification of security protocols may be considered according to protocol specifications or implementations. An ideal goal is to have a fully automated tool that verifies the security of an implemented protocol, but this goal is far from being achieved. A survey on methods and tools for verification of protocol implementations is given in [202, 276] which is beyond our discussion.

Formal methods for verification of security protocols according to their specifications can be generally divided into *model checking* and *theorem proving*. In the *model checking* approach, a finite state machine is built whose states are all possible intermediate states of protocol runs. All possible executions are then checked to satisfy a set of correctness conditions in order to find an attack on the protocol. The method concerns checking whether no state with an undesired property is reached which may indicate an attack. The correctness is concluded simply by failure in finding an attack. Model checking methods are generally more suitable for finding attacks on protocols, rather than proving their correctness. Due to possible parallel session, security protocols in general have an infinite number of states. Thus, absence of an attack in the finite model does not necessarily imply absence of an attack in infinite state. Moreover, the number of states in the finite model can be too large, and can greatly increase with increase in number of participants and performed steps. Model checking methods can provide an attack if the protocol is found not to satisfy the correctness condition. However, they do not yield a symbolic proof for the security of a protocol if an attack is not found. In the *theorem proving* approach, all possible traces of protocol runs are considered and checked to satisfy a set of correctness conditions. These methods are generally more suitable for proving the correctness, rather than finding an attack on the protocols. They may use an axiomatic (deductive) approach or an inductive approach. The axiomatic method consists of constructing a formal deductive system, called logic, whose language allows us to specify a protocol and its operations as a set of axioms. The objective is to derive the desired properties from the assumed axioms by the deduction. The inductive methods [277, 278], however, use the mathematical induction to prove that correctness conditions are valid for all traces of the protocol. Computer tools provided for theorem proving can be automatic or interactive. Automated theorem proving is a subfield of automated reasoning and mathematical logic dealing with proving mathematical theorems by computer programs. It searches for a safety proof of the protocol. If a safety proof is found, the protocol is deemed secure. However, tools for the automatic theorem proving may not terminate with a result. In interactive theorem proving, safety proofs are found by user and given to the computer for checking its correctness.

Formal methods for verification of security protocols are sometimes referred to as *symbolic models*. As mentioned earlier, symbolic models are different from the computational models. A comprehensive comparison between the symbolic and computational models is available in [202], and there are efforts for bridging the gap between these two models, mostly through the computational-soundness theorems [203]. Most symbolic formalisms are based on the Dolev-Yao model [200]. Except for a few exceptions, they offer a limited view of honesty and corruption. Principals are either honest from the be-ginning and keep their secrets to themselves or they are completely malicious and under adversarial control. This makes it hard to distinguish between the security provided by different protocols, and discerns benefits from storing long-term keys in tamper-proof modules or performing part of computations in a cryptographic coprocessor [279]. In the symbolic models, cryptographic primitives are considered as perfect blackboxes. However, they have the advantage of making it easier to build automatic verification tools, and there are numerous effective tools for symbolic protocol analysis. Formal methods analysis techniques usually rely on computer assistance, although there are exceptions such as the BAN logic [270] and its variants and the strand space model [280]. There are several automatic tools available for the symbolic analysis of security protocols, including FDR [269], Mur$\phi$ [281], Maude [282], SATMC (SAT-based model checker) [283], Brutus [284], NRL analyzer [285], PCL logic [286, 287], AAPA [288], Isabelle [289] and Isabelle/HOL [290], Athena [291], CAPSL [292], Avispa [293], Scyther [294], Scyther-proof [295], Tamarin [296], ProtoVeriPhy [297, 298], SecFuzz [299], and ProVerif [300]. The models underlying these tools differ in many aspects including protocol execution models, definitions of security properties, and the explored state space. Some tools explore all possible behaviors, whereas others explore strict subsets [301].

In contrast to the symbolic models, the adversary in the computational models does not perform predetermined actions for analyzing messages, but is modeled as an arbitrary PPT algorithm. The computational models have defined stronger adversarial power which is closer to the real execution of protocols. They allow principals to be selectively corrupted during protocol execution, for example their short-term or long-term secrets or results of the intermediate computations. Computational models provide stronger security guarantee such as perfect forward secrecy or resilience to state-reveal attacks. However, they are mostly developed just for key agreement protocols. Moreover, the proofs in the computational model are more difficult to automate [302]. Recent trends include unification of computational models and machine support. The idea stems from the fact that games are programs played with an arbitrary PPT adversary. Computer tools provided for analysis in the computational setting include Certicrypt [303], EasyCrypt [304], and CryptoVerif [305].

# 5 Conclusions

Security protocols are building blocks in secure communications. There is a wrestling between attackers and protocol designers. Including concurrent and parallel executions, there are many ways for attacking a protocol. Many protocols that have been proposed in the literature or have been implemented in practice were shown later to have security problems. In this manuscript, we reviewed foundations of security protocols, a list of standard attacks on security protocols and their implementations, security models, provable security, and different methods for security analysis of protocols. Specifically, we clarified the differences between the information-theoretic security and computational security, and the computational and symbolic models. Furthermore, we described the most important security models for AKE and PAKE protocols. With emergence of new technologies which may have different security requirements, and due to increased adversarial capabilities, there is always a need for designing new protocols. Current trends in security protocols include:

- Security protocols and models for new paradigms such as privacy-preserving, resource-constrained environments and multi-factor authentication; lightweight /ultralightweight protocols

- Security models with tighter security guarantees and decreasing the available distance between the mathematical modeling and practical security

- Finding theoretical solutions to practical problems such as leakage-resilient protocols and models

- Security protocols and models for MPC

- Post-quantum protocols and models, and quantum attacks on protocols

- Developing computer-aided methods and tools for security analysis of protocols and proofs in both symbolic and computational models

- Bridging the gap between the computational and symbolic models

## Acknowledgment

The author would like to thank Øyvind Ytrehus for helpful comments and discussions.